\documentclass[12pt,preprint]{aastex}

\shortauthors{Boroson}

\begin{document}


\title{Does the Narrow [O III] $\lambda5007$ Line Reflect the Stellar Velocity
    Dispersion in AGN?}

\author{Todd A. Boroson}
\affil{National Optical Astronomy Observatory, P.O. Box 26732, Tucson, AZ
         85726-6732}
\email{tboroson@noao.edu}

\begin{abstract}
It has been proposed that the width of the narrow [O III] $\lambda5007$
emission line can be used as a surrogate for the stellar velocity dispersion 
in active galaxies.  This proposition is tested using the SDSS EDR spectra 
of 107 low-redshift radio-quiet QSOs and
Seyfert 1 galaxies by investigating the correlation between black hole mass,
as determined from H$\beta$ FWHM and optical luminosity, and [O III] FWHM.
The correlation is real, but the scatter is large.  Without additional 
information or selection criteria, the [O III] width can predict the 
black hole mass to a factor of 5. 
\end{abstract}

\keywords{galaxies: active, bulges---quasars: emission lines}

\section{Introduction}

The correlation of nuclear black hole mass ${\rm M_{\bullet}}$ and bulge stellar 
velocity dispersion $\sigma_{\ast}$ is now well established in nearby galaxies 
\citep{tremaineetal02}. The possibility of extending the study of this relationship 
to active galaxies, using diagnostics that can be easily measured even at 
substantial redshifts, has been explored as a result of two techniques: (1) the 
use of reverberation mapping to calibrate a relation between luminosity and 
radius of the broad line region (BLR), and (2) the use of the narrow [O III]
$\lambda$5007 emission line width as a surrogate for stellar velocity dispersion.  

If the BLR is in virial equilibrium, the mass of the central black 
hole is given by M$_{\bullet} = v^2 {\rm R_{BLR}} / G$, where $v$ and ${\rm R_{BLR}}$
are the characteristic velocity and radius of the BLR.  Scale factors for converting
measured quantities to truly representative and corresponding velocities and radii
depend on the unknown kinematic structure of the BLR.  However, several studies have
shown that estimates of the black hole mass that are consistent with other methods
can be derived in this way.  \citet{gebhardtetal00} and \citet{merrittandferrarese00}
show that black hole masses derived using velocities from the H$\beta$ line width 
and radii from reverberation mapping time lags fall on the same relation in the
${\rm M_{\bullet}}$ vs $\sigma{\ast}$ diagram as those derived from spatially resolved
spectroscopy.  Admittedly, this confirmation is possible for only a small number of objects,
but the scatter appears consistent with measurement errors.  
\citet{kaspietal00} showed that a good correlation exists between BLR 
radius and the monochromatic luminosity L$_5100$ with a power law index of 0.7 for their
sample of reverberation-mapped AGNs, allowing this easily measured luminosity to be
substituted for time lags derived from arduous monitoring campaigns.  This set of 
correlations (the "photoionization method" \citep{wandeletal99}) has already 
been used in a number of studies of black hole 
masses in AGNs including \citet{merrittandferrarese00}, \citet{laor00}, \citet{lacyetal01},
\citet{shieldsetal02}, \citet{vestergaard02}, \citet{oshlacketal02}, and 
\citet{jarvisandmclure02}.

The evidence that the [O III] line width in QSOs is dominated by the gravitational 
potential on the scale of the host galaxy bulge is primarily the result of 
two studies.  \citet{nelsonandwhittle96} studied the relationship between 
the widths of narrow emission lines and near-nuclear stellar velocity dispersion
for a sample of 75 Seyfert galaxies.  They found a moderately strong correlation
between [O III] FWHM and $\sigma_{\ast}$ (r = 0.48, P(null) = 0.0038\% for 66
objects) with a slope flatter than unity due to anomolously broad [O III] lines
in objects that have powerful linear radio sources.  As a consequence of this finding,
\citet{nelson00} proposed that the [O III] FWHM could be used to
extend the ${\rm M_{\bullet}}$ vs $\sigma_{\ast}$ relation to AGNs.  His study compared
the location of 20 Seyfert galaxies and 12 QSOs in the ${\rm M_{\bullet}}$ vs [O III] FWHM
plane with the ${\rm M_{\bullet}}$ vs $\sigma{\ast}$ relation from \citet{gebhardtetal00}
(where FWHM = $2.35 \times \sigma$).
The data in his study were gathered from a number of sources in the literature, and
comprised all objects for which ${\rm M_{\bullet}}$ values had been derived from 
reverberation mapping measurements.  \citet{nelson00} found that for these AGN, the
${\rm M_{\bullet}}$ and [O III] FWHM values were strongly correlated and consistent with
the \citet{gebhardtetal00} relation, though with substantial scatter.

Because of the scatter seen by \citet{nelson00} in the ${\rm M_{\bullet}}$ 
vs [O III] FWHM relation and the heterogeneous data used in that study, it is 
useful to further investigate that relation.
If the relation is tight, it provides a mechanism for studying the evolution of
the relationship between black hole and galaxy formation processes, as the emission
lines, H$\beta$ and [O III] $\lambda$5007 can be easily observed and measured to 
substantial redshifts.  An initial attempt to compare this relationship between low
and high redshift sample using data from the literature has been published by 
\citet{shieldsetal02}.

This paper reports on an investigation of the ${\rm M_{\bullet}}$ vs [O III] FWHM
relation for AGN from a large, homogeneous data set, the Sloan Digital Sky Survey
(SDSS) Early Data Release (EDR) spectra. The goal of this study is to establish
the reality of this relation, measure its scatter, and detect any selection effects
that would compromise a comparison of low and high redshift samples.

\section{The Sample}

The QSOs included in the sample were all drawn from the SDSS EDR 
\citep{stoughtonetal02}. The
SDSS ought to provide an excellent dataset for this purpose, as 
it has uniform photometry and spectroscopy, and sufficient spectral resolution
(R $\sim$ 1800) to resolve the [O III] line \citep{yorketal00}.  Although
\citet{schneideretal02} have generated an EDR Quasar catalog, it has a 
luminosity constraint (M$_{\rm I} \leq -23$) and a line width constraint 
(FWHM $\geq$ 1000 km ${\rm s}^{-1}$).  Since the current study wants the 
largest possible
range of M$_{\bullet}$, which depends on both luminosity and line width, a new
catalog was generated from the SDSS database.  

First, spectra of all QSO candidates, having ${\rm g^*}$ magnitudes brighter than 18.0 
and redshifts less than 0.5 were downloaded from the archive. The magnitude 
limit was imposed because it was recognized that the fainter objects would
not have spectra with sufficient signal-to-noise to support the analysis. 
The redshift limit was imposed to ensure the the H$\beta$ and [O III] lines
would be in a region free of noise from strong night-sky emission lines. 
This initial
list contained 201 objects.  Visual inspection of these spectra showed that
further trimming of the list would be required to remove those objects that
were too low in signal-to-noise or which had no obvious broad line region.
This resulted in a list of 121 objects.  Two entries in this list are 
actually two independent observations of the same object, 
SDSS J032205.05+001201.4

Each spectrum was shifted to a rest wavelength scale, using the redshift
obtained from the SDSS database, and rebinned to a common format with 
1.5A ${\rm pixel}^{-1}$.  The complex Fe II emission was then measured 
and subtracted using the technique and template described in 
\citet{borosonandgreen92} (BG92).  Briefly, a template with strong, narrow Fe II
emission was constructed from a spectrum of the low redshift QSO I Zwicky 1.
Lines attributed to other ions were removed and the continuum subtracted. 
This template was then broadened and multiplied by scaling factors and
subtracted from each QSO spectrum. Different combinations of broadening 
and scaling were tried until the continuum on either side of the 
H$\beta$-[O III] complex appeared flat and featureless.  In practice, 
the initial guess for broadening was obtained by measuring the H$\beta$
width from the spectrum and assuming that the Fe II width was identical.
This produced a satisfactory result in almost all cases.

Once the best Fe II broadening and scaling were identified, a 
Fe II-subtracted spectrum was generated and the continuum was fit 
and removed between the emission lines over the range $\lambda\lambda4200-6000$. 
H$\beta$ and $\lambda5007$ widths were measured from these
normalized spectra.  These widths are not derived from fits to functions,
but are actual measured widths of the two lines at half of their maximum
intensity.  Six of the objects had no detectable (or very weak) [O III] 
emission, and so they were removed from the sample.  The remaining 115 
objects, together with H$\beta$ FWHM and $\sigma_{\rm [O III]}$ (FWHM/2.35; 
corrected for instrumental resolution; see below)
are listed in table 1.  Note that the two sets of measurements for the object
observed (and analyzed) twice, SDSS J032205.05+001201.4, differ by 5\% 
or less.

In the case of the [O III] widths, the instrumental resolution,
166 km ${\rm s}^{-1}$ (or 2.8A at $\lambda5007$), has been subtracted 
in quadrature from the measured widths.  Note that this correction only
changes the line width by as much as 20\% in 12 of the 115 objects.
 
For the H$\beta$ line widths an additional caveat is required.  In approximately
one-fourth of the spectra, a narrow H$\beta$ spike is seen on top of the
broad H$\beta$ line.  We have ignored this spike in measuring the FWHM
of the H$\beta$ line. Although such spikes include only a small fraction
of the flux in the line, they can dramatically change the FWHM value 
measured.

Recently \citet{vestergaard02} has discussed at length the correct 
method for measuring the BLR line width for the calculation of
M$_{\bullet}$.  As she points out, the appropriate BLR velocity dispersion is
that measured from the `rms' spectrum, which represents the varying part
of the line profile.  She goes on to compare the FWHM H$\beta$ values from
the rms spectra of \citet{kaspietal00} with mean and single-epoch measurements
of the same objects.  She concludes that, in general, the best way to
substitute a single-epoch spectrum for the rms spectrum is to remove the
Fe II emission, but leave the narrow component of H$\beta$ to be included
in the measurement.  The conclusion that the narrow component should not
be removed is motivated primarily by the single object PG1704+608 (3C 351),
which has a strong, narrow spike on top of a very broad H$\beta$ line.  
Measurements of the width of the broad lines in this object include 6560 km ${\rm s}^{-1}$ 
(H$\beta$; BG92), 13,000 km ${\rm s}^{-1}$ (H$\alpha$; 
\citet{eracleousandhalpern94}), 10,000 km ${\rm s}^{-1}$ (H$\beta$; 
\citet{netzeretal82}).  However, \citet{kaspietal00} quote 890 km ${\rm s}^{-1}$
for the mean spectrum and 400 km ${\rm s}^{-1}$ for the rms spectrum. Is it
really the narrow component of H$\beta$ that is varying?  

There are several reasons to think that the idea that it is the narrow
part of the line varying is in error.  First, the narrow component only
accounts for a small fraction of the line flux -- around 10\% in this
object.  Thus, to produce a change in total line flux of a factor of two, 
as is seen in the \citet{kaspietal00} data, it would need to vary by a 
factor of 20.  This is inconsistent with published spectra in 
\citet{kaspietal00} and BG92, and the SDSS EDR spectrum of this object, all
of which show the narrow H$\beta$ at similar strength relative to the
[O III] lines.  Second, the adoption of such a small characteristic 
velocity width for the BLR results in a very low black hole mass.  The
\citet{kaspietal00} value based on the rms spectrum is $7.5 \times 10^6
M\sun$.  The black hole mass based on the BG92 line width is $2.0 \times
10^9 M\sun$.  The small black hole mass would imply an unreasonably small
Eddington luminosity, resulting in an Eddington ratio of 30, while the high
black hole mass gives a more sensible Eddington ratio of 0.1.  Finally, we 
note that a drawback of the use of the rms spectrum to identify the changing
part of the line is that variations in line profiles due to seeing differences
or instrumental differences from observation to observation will cause 
residuals (see footnote 9 in \citet{kaspietal00}). The rms spectrum of
PG 1704+608 shows significant residual emission at the [O III] lines,
ten times as strong as the narrow H$\beta$ spike in that spectrum 
\citep{kaspi02}.  This
suggests that in this object, the rms spectrum is misleading as an indicator
of what part of the H$\beta$ line is varying as a response to continuum 
variations.

Table 1 also lists for each object the black hole mass, derived using the
formula M$_{\bullet} = v^2 {\rm R_{BLR}} / G$, where 
$v=\sqrt{3}/2 {\rm FWHM_{H\beta}}$, and ${\rm R_{BLR}} = 
32.9(\lambda{\rm L}_{5100}/10^{44} {\rm ergs\ s^{-1}})^{0.7}$
light days \citep{kaspietal00}. The values of L$_{5100}$ were derived from the 
${\rm r^*}$ photometry in 
the SDSS database.  Fluxes were converted to luminosities using the 
\citet{schlegeletal98} maps for correcting for galactic absorption, H$_0$ =
75 km s$^{-1}$ Mpc$^{-1}$, and q$_0$ = 0.5.  

Finally, Table 1 also lists log R, a measure of radio loudness.  The FIRST 
catalog \citep{beckeretal95} and the NVSS catalog \citep{condonetal98} were 
searched at the position of each object. All but one of the objects, 
SDSS J173348.81+585651.0, are in regions covered by one or the other radio
survey.  Any source within 30 arcseconds was declared a match, though we
note that the positional accuracy of all these surveys is good enough that
a match with a radio core ought to lie within 2 arcseconds.  Thus, we flag
those that have differences between the radio and optical centroid of more
than 2 arcseconds with a colon in Table 1.  R is the ratio of flux density
at 5 GHz to flux density at $\lambda$2500.  In computing this, a  
spectral slope of -0.3 was used to transform the total observed radio flux density
at 1.4 GHz, and an optical spectral slope of -1.0 was used to transform the 
g$^*$ magnitude.  

Figure 1 shows the black hole mass plotted against the [O III] line width
for the 115 SDSS low-redshift QSOs.  The seven objects with log R $>$ 1, the
usual criterion for radio-loud, are plotted as open circles.  The one object
unobserved in the radio is plotted as an open triangle.  The remaining 107
objects are plotted as solid squares.  

\section{Discussion}

A correlation between M$_{\bullet}$ and [O III] width is evident in Figure 1, 
though the scatter is large.  For the following fitting and statistical analysis 
the two observations of SDSS J032205.05+001201.4 have been combined by averaging 
their measurements and the radio-loud objects have been removed, resulting in a 
sample of 107 objects.  The correlation coefficient is r = 0.44 
(P = $6 \times 10^{-6}$) for the 107 objects that are not radio-loud.  
Two lines through the points are shown in 
Figure 1.  The solid line is a fit to the radio-quiet points, using the
least-squares bisector \citep{isobeetal90}, which is plausible for situations
where both variables have large uncertainties.  This is the same type of 
fit used by \citet{nelson00}.  For an equation of the form
log(M$_{\bullet}/M_\sun) = \alpha\ + \beta\ {\rm log}(\sigma/\sigma_0)$, the fit
has coefficients $\alpha$ = $8.05\pm0.07$ and $\beta$ = $3.59\pm0.47$ where
$\sigma_0 = 200 {\rm km s^{-1}}$.  The dashed line is that 
derived by \citet{tremaineetal02} who find $\alpha$ = $8.13\pm0.06$ and $\beta$ = 
$4.02\pm0.32$ 
from a sample of 31 nearby galaxies.  For comparison, the sample of \citet{nelson00},
analyzed in the same way has an identical correlation coefficient of r = 0.44, 
and coefficients of $\alpha$ = 6.54 and $\beta$ = 3.50 for the 28 radio-quiet objects.

One measure of the scatter is the standard deviation of the points from the bisector fit in
log M$_{\bullet}$.  This is the accuracy with which $\sigma_{\rm [O III]}$ could be
used to calculate M$_{\bullet}$.  For the 107 radio-quiet low-redshift objects, the standard
deviation is 0.67 in log M$_{\bullet}$, or a factor of a little less than 5.  
Alternatively, one can compare the fit to the data with the \citet{tremaineetal02}
line, which represents the relationship that we believe we are modeling.  Clearly,
both the slope and intercept (at $\sigma_0 = 200 {\rm\ km\ s^{-1}}$) are consistent
between the two fits.  This is interesting not only as confirmation that
the two approaches are demonstrating the same physical relationship, but also because
the scatter of the AGN points around the non-AGN fit is close to symmetric. 

The fact that the scatter is larger than the non-AGN sample (\citet{tremaineetal02} 
quote intrinsic dispersion in log M$_{\bullet}$ of 0.3 or less) is not surprising.
The scatter is not primarily due to measurement errors, as these are probably no
more than about 10\% for the line widths and a few percent for the magnitudes.
However, both H$\beta$ FWHM and optical brightness vary in most AGNs, and these 
will produce errors in M$_{\bullet}$.  In addition, luminosity is being used to 
predict ${\rm R_{BLR}}$, and this relation has significant scatter \citep{kaspietal00}.
On the abscissa, [O III] FWHM is not a perfect predictor of stellar velocity
dispersion \citep{nelsonandwhittle96}, but shows scatter of about 0.2 in the log
around a ratio of unity. \citet{nelsonandwhittle96} explored this relationship
in detail, and found that several properties indicated high [O III] widths relative
to the stellar velocity dispersion, including powerful linear radio sources and
systems showing obvious signs of interaction.  We have removed the radio-loud 
objects in this study, but have no way of filtering by properties such as 
host galaxy interaction.  The fact that the scatter is approximately equal on the
two sides of the \citet{tremaineetal02} fit suggests that the explanation is not as
simple as objects with anomalously high [O III] widths.

Given that one goal of this study is to lay the groundwork for an exploration of
the M$_{\bullet}$ vs $\sigma_\ast$ relationship as a function of redshift, it is
worthwhile to investigate the extent to which the fit depends on the luminosity
range of the sample.  The 107 low redshift radio-quiet objects were divided into
two sub-samples of equal size, one with ${\rm L}_{5100} < 2.7 \times 10^{44}$ 
ergs $s^{-1}$ and one with ${\rm L}_{5100} > 2.7 \times 10^{44}$ 
ergs $s^{-1}$. Figure 2 shows these two sub-samples plotted with different
symbols and the bisector fits to the sub-samples.  The coefficients of the
fits are $\alpha$ = 8.02 and $\beta$ = $2.48\pm0.72$ for the low luminosity objects
and $\alpha$ = 8.28 and $\beta$ = $1.81\pm0.52$ for the high luminosity objects.
Thus, there is a tendency for the slope to flatten in samples restricted to a
smaller range of luminosity, particularly for high luminosity.  This is not 
surprising in that the dependence of M$_\bullet$ on luminosity results in 
a dividing line between low and high luminosity samples that is flatter than
the relation itself.  This, combined with the large scatter, results in a flatter
fit.  For comparison with other samples, it
certainly seems advisable to maintain the largest possible range of luminosity.

\acknowledgments

I am grateful to Mike Brotherton and Richard Green for helpful 
conversations.

Funding for the creation and distribution of the SDSS Archive has been 
provided by the Alfred P. Sloan Foundation, the Participating Institutions, 
the National Aeronautics and Space Administration, the National Science 
Foundation, the U.S. Department of Energy, the Japanese Monbukagakusho, 
and the Max Planck Society. The SDSS Web site is http://www.sdss.org/.

This research has made use of the NASA/IPAC Extragalactic Database (NED) which 
is operated by the Jet Propulsion Laboratory, California Institute of Technology,
under contract with the National Aeronautics and Space Administration.




\clearpage

\begin{deluxetable}{lrcccc}
\tablewidth{0pt}
\tablecaption{Line Widths and Black Hole Masses of SDSS Sample}
\tablehead{
\colhead{SDSS J}           & \colhead{z}      &
\colhead{$\rm{FWHM}_{H\beta}$}    & \colhead{$\sigma_{\rm{[O III]}}$}  &
\colhead{log $M_{\bullet}$}       & \colhead{log R} \\
\colhead{}                 & \colhead{} & 
\colhead{km $s^{-1}$}      & \colhead{km $s^{-1}$} &
\colhead{$M_\sun$}}

\startdata
000011.97+000225.1 & 0.4790 & \phn2290 &   290 &  8.17 & not detected\\
000710.02+005329.0 & 0.3164 & 11158 &   271 &  9.56 & 0.27\\
001327.31+005232.0 & 0.3626 &  \phn1549 &   183 &  7.71 & not detected\\
001903.17+000659.1 & 0.0726 &  \phn3283 &    \phn90 &  8.10 & not detected\\
002444.11+003221.4 & 0.4004 &  \phn9159 &   180 &  9.56 & 1.42:\\
003238.20$-$010035.3 & 0.0919 &  \phn1919 &    \phn73 &  7.10 & not detected\\
003431.74$-$001312.7 & 0.3811 &  \phn1197 &   223 &  7.39 & not detected\\
003723.50+000812.6 & 0.2518 &  \phn2598 &   196 &  7.85 & not detected\\
003847.98+003457.5 & 0.0806 &  \phn7141 &   161 &  8.33 & 0.65\\
010342.73+002537.3 & 0.3938 &  \phn2123 &   180 &  8.07 & not detected\\

010939.02+005950.4 & 0.0929 &  \phn2765 &   144 &  7.63 & 0.29\\
011254.92+000313.0 & 0.2385 &  \phn3135 &   177 &  8.04 & not detected\\
011448.68$-$002946.1 & 0.0338 &  \phn3450 &   109 &  7.61 & 0.62:\\
011703.58+000027.4 & 0.0456 &  \phn2685 &   136 &  7.57 & -0.36\\
011929.06$-$000839.8 & 0.0901 &   \phn\phn666 &   202 &  6.17 & not detected\\
012159.82$-$010224.5 & 0.0544 &  \phn4097 &   194 &  7.97 & 0.26\\
013418.19+001536.7 & 0.3989 &  \phn4085 &   290 &  8.85 & not detected\\
013521.68$-$004402.1 & 0.0984 &  \phn1555 &   290 &  7.21 & 0.44\\
013527.85$-$004447.9 & 0.0805 &  \phn3314 &   124 &  7.93 & not detected\\
014017.07$-$005003.0 & 0.3346 &  \phn4789 &   164 &  8.98 & not detected\\

014238.48+000514.7 & 0.1458 &  \phn2783 &   121 &  7.68 & not detected\\
014644.82$-$004043.1 & 0.0827 &   \phn\phn969 &    87 &  6.80 & 0.54\\
015910.06+010514.5 & 0.2174 &  \phn3326 &   269 &  8.15 & not detected\\
015950.24+002340.9 & 0.1626 &  \phn1956 &   292 &  7.91 & 1.04\\
020615.99$-$001729.2 & 0.0427 &  \phn6887 &   169 &  8.45 & 0.06\\
021359.79+004226.8 & 0.1821 &  \phn4592 &   207 &  8.28 & 0.97\\
023335.38$-$010744.7 & 0.3679 &  \phn2444 &   247 &  8.00 & not detected\\
024651.91$-$005931.0 & 0.4681 &  \phn1426 &    \phn77 &  7.96 & 0.41:\\
025007.03+002525.4 & 0.1977 &  \phn2086 &   112 &  7.56 & not detected\\
025505.67+002523.0 & 0.3541 &  \phn4715 &   229 &  8.63 & not detected\\

025646.97+011349.4 & 0.1766 &  \phn2512 &   226 &  7.73 & not detected\\
030124.26+011022.9 & 0.0715 &  \phn3073 &   183 &  7.57 & not detected\\
030144.20+011530.9 & 0.0747 &  \phn4221 &   215 &  7.93 & not detected\\
030417.78+002827.3 & 0.0445 &  \phn1173 &    \phn94 &  6.58 & not detected\\
030639.58+000343.1 & 0.1073 &  \phn2567 &   196 &  7.75 & 0.66\\
031027.83$-$004950.8 & 0.0804 &  \phn3073 &    \phn66 &  7.75 & not detected\\
031427.47$-$011152.4 & 0.3869 &  \phn1660 &   188 &  7.77 & not detected\\
032205.05+001201.4 & 0.4719 &  \phn3604 &   247 &  8.76 & not detected\\
032205.05+001201.4 & 0.4717 &  \phn3789 &   255 &  8.80 & not detected\\
032213.90+005513.5 & 0.1851 &  \phn2339 &   118 &  8.06 & not detected\\

032337.65+003555.6 & 0.2154 &  \phn1395 &   194 &  7.28 & not detected\\
032559.97+000800.7 & 0.3606 & 12380 &   175 &  9.57 & not detected\\
032729.89$-$005958.4 & 0.1340 &  \phn6381 &   164 &  8.61 & not detected\\
034226.50$-$000427.1 & 0.3762 &  \phn2573 &   239 &  8.18 & not detected\\
101044.51+004331.2 & 0.1780 &  \phn6221 &   177 &  8.85 & -0.29\\
101314.87$-$005233.6 & 0.2759 &  \phn1506 &   368 &  7.44 & not detected\\
102448.57+003537.9 & 0.0954 &  \phn1481 &    \phn90 &  7.04 & not detected\\
102936.10$-$010201.0 & 0.1398 &  \phn5894 &   199 &  8.35 & not detected\\
103457.29$-$010209.1 & 0.3280 &  \phn1216 &   124 &  7.40 & not detected\\
104230.14+010223.7 & 0.1155 &   \phn\phn673 &   166 &  6.53 & not detected\\

104332.88+010108.9 & 0.0719 &  \phn2499 &   109 &  7.55 & not detected\\
105706.94$-$004145.1 & 0.1876 &  \phn1784 &   169 &  7.50 & not detected\\
110057.71$-$005304.6 & 0.3776 &  \phn4851 &   166 &  8.71 & not detected\\
105935.76$-$000551.3 & 0.2825 &  \phn2296 &   150 &  7.82 & not detected\\
113541.21+002235.3 & 0.1753 &   \phn\phn994 &   103 &  6.88 & not detected\\
113909.66$-$001608.7 & 0.1351 &  \phn3598 &   115 &  7.79 & not detected\\
114335.37$-$002942.4 & 0.1715 &  \phn7048 &   115 &  8.54 & not detected\\
113909.66$-$001608.7 & 0.1351 &  \phn3752 &    \phn97 &  7.83 & not detected\\
113923.66+002301.6 & 0.4721 & 15559 &   329 &  9.92 & not detected\\
115235.00$-$000542.7 & 0.1288 &  \phn3894 &    \phn90 &  7.81 & not detected\\

115758.73$-$002220.9 & 0.2598 &  \phn4635 &   141 &  8.65 & not detected\\
120014.08$-$004638.7 & 0.1793 &  \phn2290 &   161 &  7.62 & 1.80:\\
122432.41$-$002731.5 & 0.1571 &  \phn1160 &   136 &  6.85 & not detected\\
124324.22+010028.1 & 0.0897 &  \phn5604 &   121 &  8.02 & not detected\\
124623.00+002839.9 & 0.0884 &  \phn2518 &   115 &  7.40 & 0.78\\
130023.22$-$005429.8 & 0.1222 &   \phn\phn839 &   118 &  6.45 & not detected\\
130713.25$-$003601.7 & 0.1700 &  \phn1537 &   158 &  7.28 & not detected\\
130756.58+010709.6 & 0.2754 &  \phn3913 &   290 &  8.43 & not detected\\
131108.48+003151.8 & 0.4293 &  \phn1401 &   242 &  7.61 & not detected\\
132135.33$-$001305.8 & 0.0822 &  \phn4289 &   147 &  8.04 & 0.34\\

134044.52$-$004516.7 & 0.3844 &  \phn3215 &   292 &  8.42 & not detected\\
134113.94$-$005315.0 & 0.2373 &  \phn2030 &   239 &  7.67 & 1.17\\
134251.61$-$005345.4 & 0.3259 &  \phn3166 &   492 &  8.43 & not detected\\
134351.07+000434.7 & 0.0736 &  \phn1672 &    \phn87 &  6.87 & not detected\\
134452.91+000520.2 & 0.0871 &  \phn2000 &   166 &  7.18 & not detected\\
134459.45$-$001559.5 & 0.2448 &  \phn2024 &   121 &  7.65 & not detected\\
135943.14$-$003424.6 & 0.1630 &  \phn2105 &   150 &  7.50 & not detected\\
143704.12+000705.0 & 0.1403 &  \phn1641 &    \phn84 &  7.04 & not detected\\
143847.54$-$000805.5 & 0.1040 &  \phn4123 &   118 &  7.83 & not detected\\
144930.49$-$004746.4 & 0.2532 &  \phn8239 &   175 &  8.92 & 0.91\\

144932.70+002236.3 & 0.0806 &   \phn\phn882 &    \phn83 &  6.47 & not detected\\
145631.65$-$001114.2 & 0.1325 &  \phn7542 &   118 &  8.40 & not detected\\
151722.52$-$003002.8 & 0.4450 &  \phn2085 &   172 &  8.15 & not detected\\
151723.24$-$002709.3 & 0.1218 &  \phn4376 &   188 &  8.02 & 0.89\\
151956.57+001614.6 & 0.1145 &  \phn1537 &   103 &  6.99 & not detected\\
152035.35$-$002040.1 & 0.1303 &  \phn5788 &   138 &  8.32 & not detected\\
152203.77+001128.3 & 0.2400 &  \phn3012 &   221 &  8.00 & not detected\\
152628.20$-$003809.4 & 0.1235 & \phn2203 &    \phn87 &  7.48 & 0.65:\\
154344.28$-$001452.2 & 0.3018 &  \phn2839 &   147 &  8.13 & not detected\\
165338.68+634010.6 & 0.2793 &  \phn1660 &   311 &  7.52 & not detected\\

165627.32+623226.9 & 0.1848 &  \phn5271 &   191 &  8.42 & not detected\\
165958.94+620218.1 & 0.2323 &  \phn3919 &   169 &  8.17 & not detected\\
170328.96+614109.9 & 0.0773 &  \phn5480 &   196 &  8.18 & not detected\\
170441.38+604430.4 & 0.3719 &  \phn9504 &   167 &  9.92 & 2.15\\
171049.89+652102.1 & 0.3855 &  \phn5098 &   386 &  8.72 & not detected\\
171300.69+572530.3 & 0.3603 &  \phn2660 &   352 &  8.06 & 0.65\\
171411.63+575834.0 & 0.0927 &  \phn2129 &   180 &  7.53 & not detected\\
171550.50+593548.8 & 0.0658 &  \phn5326 &   121 &  8.47 & not detected\\
171737.95+655939.3 & 0.2927 &  \phn5017 &   290 &  8.66 & not detected\\
171750.60+581514.1 & 0.3101 &  \phn4092 &   253 &  8.50 & not detected\\

171829.00+573422.4 & 0.1007 &  \phn1388 &   205 &  7.09 & not detected\\
171902.30+593715.9 & 0.1785 &  \phn2086 &   124 &  7.53 & not detected\\
172032.29+551330.2 & 0.2729 &  \phn3604 &   138 &  8.16 & not detected\\
172026.70+554024.2 & 0.3595 &  \phn2777 &   177 &  8.33 & not detected\\
172533.08+571645.6 & 0.0659 &  \phn7480 &   109 &  8.25 & not detected\\
172711.82+632241.9 & 0.2176 & 10535 &   169 &  9.18 & not detected\\
173107.87+620026.1 & 0.0687 &  \phn4863 &   164 &  8.19 & not detected\\
173348.81+585651.0 & 0.4911 &  \phn4277 &   202 &  8.71 & not observed\\
232259.99$-$005359.4 & 0.1503 &  \phn2617 &   213 &  7.85 & 0.20\\
232328.00+002032.9 & 0.1196 &  \phn4320 &   150 &  8.18 & not detected\\

233908.80$-$000637.8 & 0.4826 &  \phn6795 &   242 &  9.10 & not detected\\
234141.51$-$003806.7 & 0.3192 &  \phn1820 &   175 &  7.75 & 0.44:\\
234932.77$-$003645.9 & 0.2790 &  \phn3117 &   308 &  8.22 & 1.27:\\
235156.12$-$010913.4 & 0.1741 &  \phn5332 &   175 &  8.72 & 2.48\\
235457.10+004219.9 & 0.2705 &  \phn5721 &   440 &  8.61 & not detected\\

\enddata

\end{deluxetable}


\begin{figure}
\plotone{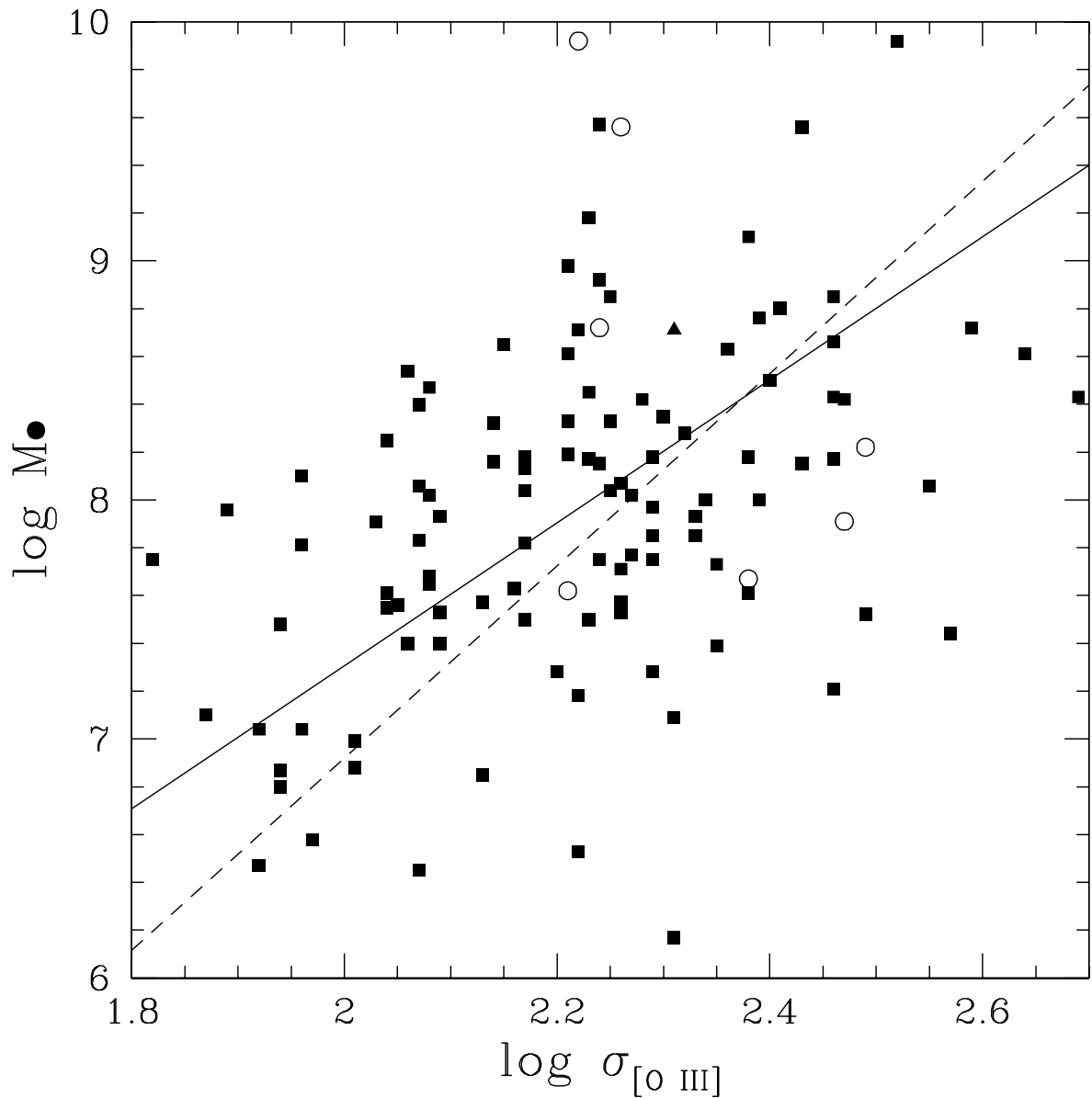}
\caption{Log of the black hole mass in solar masses plotted against log of the 
velocity dispersion in km ${\rm s^{-1}}$ from the [O III] $\lambda$5007 line. 
Radio quiet objects are shown as solid squares. Radio loud objects are shown 
as open circles. The single object unobserved 
in the radio is shown as an open triangle.  The solid line is a bisector fit 
to the radio quiet points.  The dashed line is the Tremaine etal. (2002) fit
to nearby galaxies.\label{fig1}}
\end{figure}

\clearpage 

\begin{figure}
\plotone{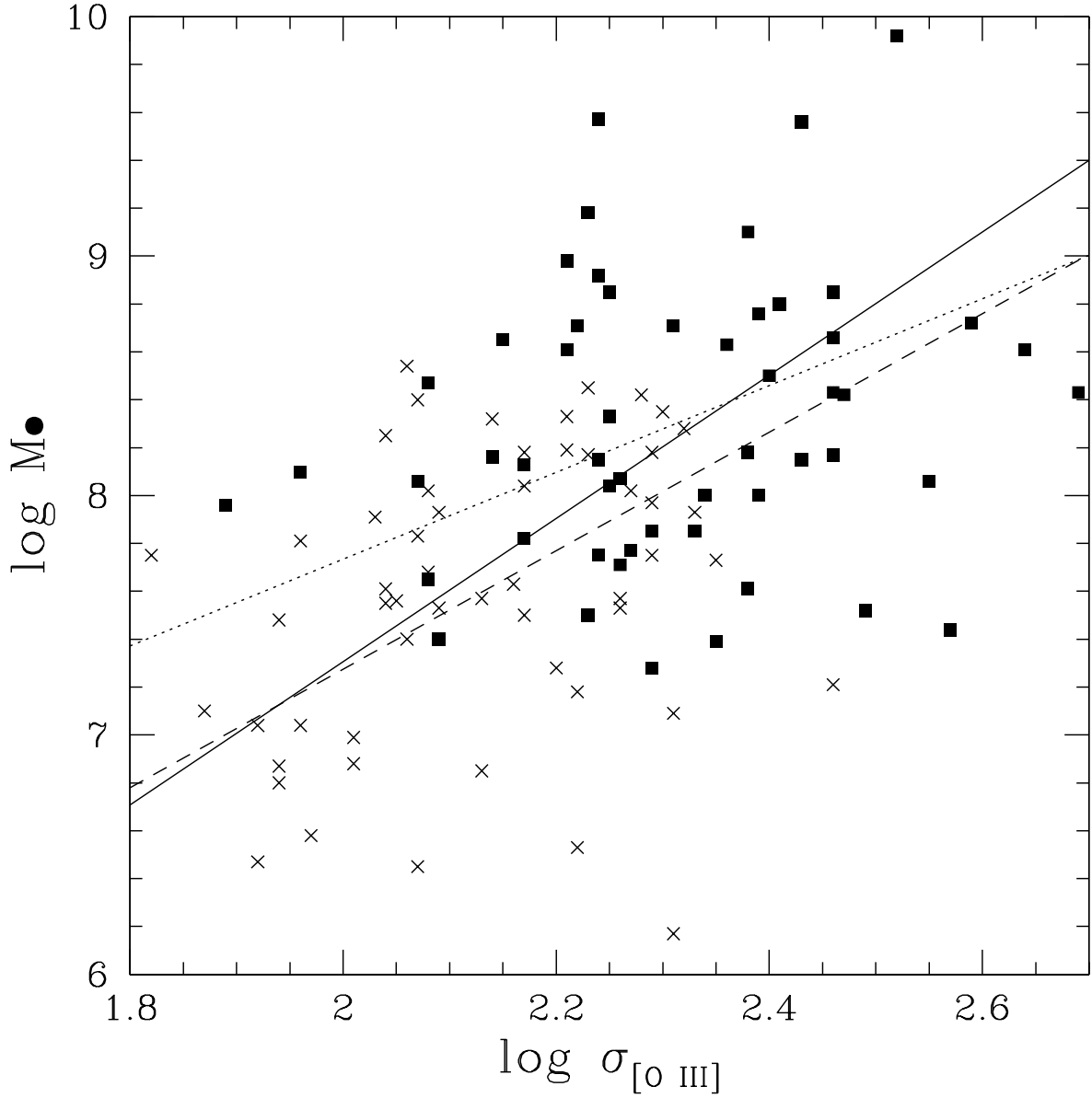}
\caption{Same as Figure 1 but with radio-quiet objects divided into low ($\times$) and
high (squares) luminosity subsamples.  Fits are to (dashed) low luminosity, (dotted) 
high luminosity and (solid) whole sample.\label{fig2}}
\end{figure}


\clearpage


\end{document}